\documentclass[acus]{JAC2000}

\usepackage{graphicx}

\begin{document}
\title{\flushright{TUAP019}\\[15pt] \centering INSERTION DEVICE CONTROLS AT THE SWISS LIGHT SOURCE}

\author{T.Korhonen, B.Kalantari \\ Paul Scherrer Institute, Villigen, Switzerland}

\maketitle

\begin{abstract}

In the initial phase of the Swiss Light Source (SLS), four insertion devices
(ID) will be installed for the first four beamlines. 
The control system for all the ID's follows, where possible, a uniform 
concept to allow a rapid installation schedule, while at the same time 
allowing for a variety in operational requirements. The components of the
control system include the gap drive system with encoders, 
temperature monitoring, beam position monitoring, power supply 
controllers for corrector magnets \cite{andreas-psc}  and a timing
module \cite{timingpaper}.
 The gap drive system requirements range from micron-level accuracy to 
driving double undulators in synchronism with each other and also with 
the other beamline elements. The local synchronism is achieved with a 
tightly coupled system having an intelligent motor controller and the 
global synchronism by extending the machine timing distribution to the 
insertion devices and the beamlines with the capability to add 
beamline-specific timing events.

\end{abstract}

\section{INTRODUCTION}

At the SLS, the first insertion devices are a wiggler for the material
science beamline, an in-vacuum minigap undulator for protein
crystallography, an elliptically polarizing undulator (two identical 
devices) and a long electromagnetic undulator (two devices.)
The core of the control system is specific for each device, but they also
have a number of components in common.
 
The first device installed was an in-vacuum undulator (U24) developed
in collaboration with Spring-8 in Japan. Its gap drive has a single
stepper motor and an absolute rotary encoder. These are controlled through
a local controller that has a GPIB bus interface. 
To integrate it with our control system, we wrote EPICS device 
support code for that controller in order to poll the status and send 
commands to drive the gap to a desired position. 
The common part of the control system and the user interface, however, 
was developed for this device first and adapted for the others.

For the two other permanent magnet devices, namely the wiggler W61 and 
a double undulator UE56, the gap controller had to be developed in-house. The
aim was to design a system that would serve as a basis for the control
system of future insertion device development.

The wiggler W61 has two motors to control the gap. It was decided to have
two motors to be able to control the parallelism (taper) of the magnet 
arrays.

The UE56 is an undulator capable of producing linearly and elliptically 
polarized light. The switching of the polarization mode is achieved by 
splitting the upper and lower magnet arrays into two halves and having 
a mechanism move them relative to each other. There are a total
of four motors: two for the gap and two for controlling the shift arrays.
Two identical IDs will be installed for the beamline and thesewhich have to
be controlled synchronously, giving us 8 axes to control.

The fourth device is an electromagnetic double undulator UE212. This ID
has no moving parts, but the fields are generated by electromagnets. For
power supplies for this device we use the standard power supplies of the
SLS machine \cite{andreas-psc}. Very little additional development was 
necessary, except for the operation of the device in several modes defined for
it. These could be implemented as additions on top of the generic SLS power
supply control software.

\section{Gap Control}

Gap control is the centerpiece of an insertion device control system. 
For the W61 and UE56, the gap control system had to be developed 
in-house. Although the devices are quite different, we strived to find
a common design that could also be used as a basis for future developments.

The gap control consists of a drive system to 
move the magnet arrays and a position measurement system to monitor the 
distance (gap) between them. 
The positions are typically measured by encoders. The philosophy was to have 
a gap measuring device as close as possible to the real mechanical gap. 
To have a high precision over a quite large range, we decided to use 
linear incremental encoders manufactured by Heidenhain \cite{heidenhain}. 
The selected encoders (ULS300) have high precision absolute reference marks. 
This allows high reproducibility of the gap setting.

For the motor controller, we selected the Oregon Microsystems 
OMS58 motor controller card \cite {oms}. This card has software support 
from the EPICS community and is already in wide use at SLS beamline 
controls. One big advantage is also that this card has support for 
both servo and stepper  motors, with the same software 
interface, so a large part of the software for both the
servo and stepper motor-based systems could be reused.

The devices had a number of different requirements: the ability to 
drive to a fixed setting with open loop control only and real time 
position hold feedback. In all cases, several motors have to be moved 
synchronously.

For the development of the gap drive we decided to build an ``ID test stand'',
a mechanical model of an ID with which we could test and debug the control
system before having the actual devices. The test system included only the gap
drive and the linear encoders; it turned out to be an extremely valuable
development tool, especially because the time between ID delivery to PSI
adn installation was short and the time for control system development 
very limited. With the test stand the control could be developed without 
having the actual device. Without the test stand it would have been almost 
impossible to meet the construction schedule.

For the W61, the gap accuracy requirement is not that strict and the movement
is only in the gap direction, so it was sufficient to have stepper motors and
an open loop control with capability to monitor the linear encoders and, if
necessary, do a position correction.
The system has two motors that have to run synchronously to keep the taper, 
i.e., the inclination angle of the magnet arrays, close to zero. To have a
protection in the low level we built a controller with a PLC to act as an 
interface between the power drive system and the control system. 
The PLC monitors that both motors are running and that the taper does not 
get too large. It stops the drives if there is a problem.
The foreseen operation mode for the wiggler is to be driven to a gap
value and left there for a long period of data taking during the experiment. 
For this reason, there were no requirements of closed loop control; 
an open-loop control with stepper motors is adequate. The system can
be controlled with high resolution by correcting the position with the
help of the linear encoders.

The UE56 has more demanding requirements. The device has the ability
to change the light polarization from linear to elliptical by 
shifting the upper and lower magnet arrays relative to each other. The 
large magnetic forces push mechanically the gap open when the shift 
arrays are moved. The difference is measured to be about 300 microns, 
which is not acceptable for operation. The system required closed 
loop position control, with four axes. 
The drive system uses AC servo motors. The conventional way would be to use 
rotary encoders on the motors for the position feedback, but to ascertain the 
required high precision, we decided to do the feedback directly from the 
linear encoders. This is more difficult to tune, because the whole system
is included in the feedback loop and the components cannot be individually
tuned. However, this simplifies the additional software because the regulation
is directly based on the gap and no corrections for the effects of the 
mechanical elements in between (like backlash, bending, creep) are required.

Like for W61, a PLC system for low-level local control was developed.
This time an external company implemented our specifications.

Two identical devices will be installed for the beamline. The reasons for
having two devices are first to have a larger photon flux, and to have
a fast switchable polarisation selection available. The switching is achieved
by steering the beam horizontally a little apart through the undulators,
putting them into different polarization modes and then having a mechanical
beam chopper to select one of the two photon beams. In this mode, the two
undulators have to be controlled as a single device.

We have achieved 1 micron precision and repeatability for the UE56 gap drive. 
The polarization mode can be changed dynamically and the feedback keeps the 
gap value constant regardless of the highly nonlinear magnetic forces from 
the magnet array shifts.

\section{Additional systems}

For the operation of the insertion devices, additional monitoring
systems are necessary to guarantee the safety of the operation and
to optimize the performance.

\subsection{Beam Position Monitors (BPM)}

For monitoring the beam position close to the insertion device, we use
a separate beam position monitor system, consisting of pickups, a
BPM processing module (Bergoz) \cite {bergoz} and a (SLS standard) 
ADC to record the positions. The main purpose of these BPMs is to 
provide signals to an external interlock system that could dump the 
beam if the beam orbit had a large offset or a large angle at the 
insertion device. These BPMs have proven very valuable to 
monitor the effect of the ID on the beam because they are situated 
close to the ID and give direct information about
the beam orbit at that point. Further study is necessary to fully
understand the behaviour and calibration of these BPMs.

\subsection{Beam Loss Monitors}

Beam loss monitors have been installed near the insertion
devices. Especially critical is the U24 in-vacuum undulator. 
For that there are large area scintillators.
The scintillators are very sensitive and have a fast response. They
have been proven to be very useful for commissioning.
The output from the beam loss monitors is pulses withthe frequency giving
the loss rate. The pulses are read out with a (SIS \cite{sis}) multichannel 
scaler into EPICS channels.

\subsection{Temperature measurement}

Monitoring of the vacuum chamber temperature is a safety measure
against heating when the orbit is bad, or, as in the case of the
in-vacuum undulator also serves as an indicator of possible problems in 
the cooling system. 
In this device, the control program constantly monitors the temperature and
if it rises over a specified threshold, the gap is automatically opened.
The temperatures are measured with thermocouples and Greenspring 
\cite {greenspring} thermocouple IP (Industry Pack) cards.
For U24 we can also monitor the cooling system through a serial port 
interface.

\subsection{Correctors}

Each insertion device has horizontal and vertical corrector magnets
placed close to the device upstream and downstream. These correctors
are used to minimize the effect of the residual kicks from the ID
at different settings. The corrector values for each gap setting
follow breakpoint tables that have been measured during commissioning 
of each device.
At the time of writing, the detailed implementation of these lookup
tables for the UE56 is still partially open because there are two
parameters: the gap and the polarization mode (magnet array shift.)
Both of these have an effect of the orbit. We need two- (or 
perhaps even three-) dimensional lookup tables to fully implement the
automatic correction. 
 
\subsection{Synchronization}

In the future the insertion devices should be operated 
in synchronism with the other beamline components. In most cases this 
can be achieved simply with EPICS Channel Access.
A method for tighter synchronization is foreseen with an extension 
of the machine event distribution system to the beamlines.
Although the main interest of tight synchronization is in the experimental
stations, the extension is done in the ID system by putting an event
generator in the ID crate. In this way, additional events that are specific to each beamline can be generated by accessing the event generator.
An example would be a sequence of movements that could be pre-loaded to
the beamline control IOCs and synchronously triggered by sending the
events from the ID control system crate.

\section{The Software}

The control system is implemented in EPICS, using mostly readily available
driver and device support modules plus standard records. So far, all the 
functionality could be achieved without having
to modify any of the low-level components (record types etc.), except, of 
course, writing device support for cards that do not have it already.

\subsection{User Interface}

Although the devices are very different, much effort has been taken to
make them look similar to the operator and to hide the different
levels of complexity behind a common operating interface. Basically, only
a minimum of information is normally presented. The operator normally
can only switch on or off the drives and to set the gap to the desired value.
For instance, before turning on the power the devices go internally 
through a series of checks and settings (the gap setpoint is synchronised
with the actual gap value, for instance) before the motor power is turned
on.

At the time of writing the devices are still considered to be in 
commissioning and the user interface is rather rudimentary.

\subsection{Integration to the beamline}

The devices have been operated as independent units without any direct
connection to the corresponding beamline. In the future, however, it will 
become increasingly important that the beamline components are aware of each 
other and can be controlled as one system. For this, it is required to agree
on the method and rules of communication between the component control systems.
The fact that all the beamline components are controlled with EPICS eases 
the integration. 

The regular operation when a device is set to a certain gap/energy value is 
straightforward. The system will getmore complicated when we want to do 
synchronised scans, i.e., operate the beamline monochromator in sync with 
the insertion device. Since the monochromator and the gap control 
are fairly slow devices, the synchronization is easily achieved with EPICS
Channel Access over the network.

\section{CONCLUSIONS}

We have initially commissioned all the four types of planned insertion devices
at SLS. Although work still remains to be done to finalize the operator 
interface and debug and calibrate the components, the concept has proven to
work quite well and fulfills all the requirements.
Two big factors that contributed to its successful development were the 
strong support and availability of software from the EPICS community, and 
the fact that we were able to standardize our control hardware to a large extent and could reuse the elements and we also had an extensive in-house 
expertise with the components.
One big factor was having built a ``test stand'', a mechanical model of an
insertion device, with the ability to simulate (nonlinear) magnetic forces.
Using the test stand we could develop the gap drive system before the 
devices were available for testing; this gave us almost one year of lead time 
before the delivery of the device. When the real device arrived, 
testing and commissioning could proceed smoothly without big surprises.

\end{document}